\documentclass[11pt,a4paper]{article}
\usepackage[T1]{fontenc}
\usepackage[utf8]{inputenc}
\usepackage{authblk}

\usepackage{geometry}
\usepackage{amsfonts,amsmath,amssymb,amsthm,nicefrac}
\usepackage{ulem}
\usepackage{cancel}
\usepackage{yhmath}
\usepackage{arcs}
\usepackage{graphicx}
\usepackage{url}
\usepackage{subcaption}
\usepackage{url}
\usepackage{caption}
\usepackage{subcaption}

\DeclareMathOperator{\arcoth}{arcoth} 

\title{\bf Comparing classicality of qutrits from Hilbert-Schmidt, Bures and Bogoliubov-Kubo-Mori ensembles}
\author[1,2,3]{Arsen Khvedelidze}
\author[3]{Astghik Torosyan}

\affil[1]{A. Razmadze Mathematical Institute, Iv. Javakhishvili Tbilisi State University, Tbilisi, Georgia}
\affil[2]{Institute of Quantum Physics and Engineering Technologies, Georgian Technical University, Tbilisi, Georgia}
\affil[3]{Laboratory of Information Technologies, Joint Institute for Nuclear Research, Dubna, Russia}

\date{ }

\begin{document}

\maketitle

\begin{abstract}
In the report we analyze the indicator/measure of classicality of quantum states defined as the probability to find a state with a positive Wigner function within a unitary invariant random ensemble. The indicators of classicality of three ensembles associated with the Hilbert-Schmidt, Bures and Bogoliubov-Kubo-Mori metrics on the space of quantum states of 3-level system are computed. Their dependence on a moduli parameter of the Wigner function is studied for all strata of a qutrit state space stratified in accordance with the unitary group action.
\end{abstract} 

\tableofcontents

\newpage 


\section{Introduction}

It is natural to expect that some states of a quantum system are more ``quantum'' than the others. To transform this intuitive thought into a qualitative concept, we use the conventional statistical interpretation of quantum mechanics. The quasiprobability distribution functions will be regarded as a source of information about the classicality/quantumness of a state.  Our consideration is based on the ideas borrowed from the geometric probability theory \cite{KlainRota1997} and a commonly accepted opinion that if quasiprobability functions attain negative values, then it is a certain sign of quantum nature (see \cite{Wigner1932}-\cite{Feynman1987} and \cite{FerrieMorrisEmerson2010} with references therein). This observation allows one to specify the notion of ``classical states à la Wigner'' as the states whose Wigner function is positive semidefinite everywhere in the phase space. Based on this definition, several measures of classicality/quantumness have been constructed  \cite{Hillery1987}-\cite{AKA2021}. When dealing with an ensemble of random states, the probability to find a ``classical state'' among the members of an ensemble is an example of these kind of measures \cite{AKhT2020}-\cite{AKR2021}. 

In the present article, after the introduction of classicality indicator $\mathcal{Q}$ as the geometric probability, we will compute it for a 3-level quantum system, a qutrit. We will compare the characteristics of classicality of qutrits from three random ensembles: the Hilbert-Schmidt and two other ensembles, associated  with the monotone Riemannian metrics \--- the Bures and the Bogoliubov-Kubo-Mori metrics (cf. \cite{MorozovaChensov1990}-\cite{HiaiKosakiPetzRuskai2013}). To make the presentation self-consistent, in the next sections necessary notions and definitions related to these random ensembles and the Wigner function of a finite-dimensional quantum system will prelude calculations of the corresponding probabilities. Calculating the probabilities for different varieties of states, we analyze the dependence of the classicality measure on the moduli parameter of a qutrit Wigner function.

\section{Unitary invariant ensembles of qudits}

\label{sec:Random}

Let us consider a qudit \--- a quantum system associated with an $N\--$dimensional Hilbert space. The quantum state space $\mathfrak{P}_N$ of an $N\--$level qudit is defined as: 
\begin{equation}
\label{eq:StateSpace}
\mathfrak{P}_N =\{\, \varrho \in M_N(\mathbb{C}) \ |\ \varrho=\varrho^\dagger\,,\quad  \varrho \geq 0\,,  \quad \mbox{tr}\left( \varrho \right) = 1  \, \}\,.
\end{equation}
The unitary $U(N)$ automorphism of the Hilbert space of an $N\--$level quantum system induces the adjoint $SU(N)$ transformations of density matrices $\varrho \in \mathfrak{P}_N$\,: 
\begin{equation}
\label{eq:UP}
    g\cdot \varrho = g \varrho g^\dagger\,, \qquad  g\in SU(N)\,.
\end{equation}
For a closed system it is assumed that the probability density function of the corresponding  ensemble of $N\--$ dimensional qudits is invariant under (\ref{eq:UP}):
\begin{equation}
\label{eq:InvPDF}
    P(\varrho)= P(g\varrho g^\dagger)\,, \qquad  \forall\ g \in SU(N)\,.
\end{equation}
Further in the report three ensembles of random states respecting this unitary symmetry will be used for evaluation of the measure of classicality. Namely, we will consider the unitary invariant ensembles associated with the following Riemannian metrics on state space:
\begin{itemize}
\item[\--] the Hilbert-Schmidt metric $\mathrm{g_{{}_\mathrm{HS}}}$\,; 
\item[\--] the Bures metric $\mathrm{g_{{}_\mathrm{B}}}$\,;
\item[\--] the Bogoliubov-Kubo-Mori metric $\mathrm{g_{{}_\mathrm{BKM}}}$\,.
\end{itemize}
Before dealing with a specific ensemble, it is worth drawing attention to a common property of each of these ensembles emerging due to $SU(N)$ invariance (\ref{eq:InvPDF}). 


\paragraph{Stratification and factorization of probability distribution on $\mathfrak{P}_N$\,.}
The invariance property (\ref{eq:InvPDF}) leads to a certain factorization of the probability distribution functions $ P(\varrho)$ into two factors, one depending on $SU(N)\--$invariants solely, and the other being a universal function of the ``angular variables''.
Moreover, the structure  of this factorization is universal for all  states whose unitary orbits are  characterised by the same isotropy group, $H_\alpha \subset SU(N)$\,,  i.e., belong to a class with the same ``orbit type'' 
\footnote{
Subgroup $H_x \subset SU(N)$ is the \textit{isotropy group (stabilizer)} of point $x\in \mathfrak{P}_N$ and is defined as 
\[
H_x =\{\, g\in SU(N)\ | \ g\cdot x =x\, \}\,.
\]
If the conjugacy class of $H$ is denoted  by $[H]$\,, then we say that \textit{the type of the orbit is $[H]$}, if the stabilizer $H_x$ of some/any point $x$ in the orbit belongs to $[H]\,.$ 
}\,.
Isotropy groups $H_\varrho$ of any point $\varrho  \in \mathfrak{P}_N$ are determined by the algebraic degeneracy of the spectrum of $\varrho$ and are in one-to-one correspondence with the Young diagrams of all possible decompositions of $N$ into non-negative integers. Hence,  we associate the given partition of $N$ with the \textit{stratum}   $\mathfrak{P}_{[H_\alpha]}\,,$ defined as the set of all points of $\mathfrak{P}_N\,,$ whose stabilizer is conjugate to subgroup $H_\alpha$\,:
\begin{equation}
   \mathfrak{P}_{[H_\alpha]}: =\big\{\, x \in \ \mathfrak{P}_N|\   H_x \mbox{~is~conjugate~to}\  H_\alpha \, \big\} \,,
\end{equation}
where  $ \ \alpha = 1, 2, \dots, p(N)$\,. \footnote{The partition function $p(N)$  gives a number of possible partitions of a non-negative integer  ${N}$ into natural numbers.
}
The union of $\mathfrak{P}_{[H_\alpha]}$ results in the state space $\mathfrak{P}_{N}$\,:
\begin{equation}
\label{eq:OrbitDec}
\mathfrak{P}_N=\bigcup_{\mbox{orbit types}}{\mathfrak{P}}_{[H_\alpha]}\,,
\end{equation}
with each component of the decomposition 
(\ref{eq:OrbitDec}) consisting  of density matrices with a fixed  algebraic degeneracy,
\begin{equation}
\label{eq:stratumDeg}
\mathfrak{P}_{[H_\alpha]}=
 \bigcup_{\omega \in S_s }{\mathfrak{P}_{k_{\omega(1)},
 k_{\omega(2)},
 \dots, k_{\omega(s)}}}\,.  \end{equation}
In (\ref{eq:stratumDeg}) $S_s$ is a symmetric group acting on a given partition of $N$ into $s$ natural numbers  $k_1, k_2, \dots, k_s\,.$ Algebraically, $\mathfrak{P}_{k_1, k_2, \dots, k_s}$ being a set of states with a fixed degeneracy is defined via the characteristic polynomial of a density matrix:
\footnote{Note that in (\ref{eq:DegSet}) the condition of summing up the degrees of degeneracy to $N$ means that only the maximal rank states are considered.}
\begin{equation}
\label{eq:DegSet}
 \mathfrak{P}_{k_1, k_2, \dots, k_s} = \{\, \varrho\in\mathfrak{P}_N\,, k_i \in \mathbb{Z}_+\, |\, 
\det(\varrho-\lambda)=\prod_{i=1}^s (r_i-\lambda)^{k_i}\,, \quad   \sum_{i=1}^s k_i= N \, \}\,.
\end{equation}
Geometrically, the set $\mathfrak{P}_{k_1, k_2, \dots, k_s}$ with $k_1=k_2=\cdots=k_N=1$ represents the interior of an $(N-1)\--$dimensional simplex $C_{N-1}$ of eigenvalues:
\begin{equation}
\label{eq:NorderedSim}
 C_{N-1} := \{\, \boldsymbol{r} \in \mathbb{R}^N \, \biggl| \, 
\sum_{i=1}^{N} r_i = 1\,, \quad 1\geq r_1\geq r_2 \geq \dots \geq r_{N-1}\geq r_N \geq 0 \, \}\,, 
\end{equation}
while for all other admissible tuples $\boldsymbol{k}=(k_1, k_2, \dots, k_s )$ each $\mathfrak{P}_{k_1, k_2, \dots, k_s}$ represents the union of the faces and edges of the $(N-1)\--$simplex parameterized by the barycentric coordinates of the following kind: 
\begin{equation}
\label{eq:spec}
\boldsymbol{r}^{\downarrow}(\varrho)=\{r_1 \overbrace{(1, \dots, 1)}^{k_1}\,;\, r_2\overbrace{(1, \dots, 1)}^{k_2}\,;\, \dots \,;\, r_s\overbrace{(1, \dots, 1)}^{k_s}\}\,.
\end{equation}

Now, bearing in mind the above described stratification of $\mathfrak{P}_N\,,$ it is easy to show the factorization of $SU(N)\--$invariant measures. Indeed, one can be convinced that the Singular Value Decomposition (SVD) of the density matrix from a stratum $\mathfrak{P}_{[H_\alpha]}$ with spectrum of the form (\ref{eq:spec}): 
\begin{equation}
\label{eq:SVD}
\varrho=U\mathrm{diag}\left(r_1, r_2 , \dots r_s\right) U^\dagger\,,  \qquad U\in SU(N)/H_\alpha\,,
\end{equation} 
reveals the following factorization of the invariant probability distribution (\ref{eq:InvPDF}):
\begin{equation}
   \mathrm{P}(\varrho)=P(r_1,\dots, r_s)\, \mathrm{d}r_1\wedge \cdots \wedge \mathrm{d}r_N\wedge \mathrm{d}\mu_{U(N)/H}\,,
\end{equation}
where the first factor $P(r_1,\dots, r_s)$ represents a measure on
subset $\mathfrak{P}_{k_1,k_2,\dots,k_s}$ of the simplex $\mathcal{C}_{N-1}$, while the second factor is the measure on coset 
$U(N)/H$\,. 

After a preliminary exposition of this generic property of unitary invariant ensembles, we will now specify the form of the distribution $P(r_1,\dots, r_N)$ for the  Hilbert-Schmidt metric and for an important class of the monotone metrics.


\paragraph{The Hilbert-Schmidt ensemble of qudits.}
Let us consider the metric corresponding to the distance between two infinitesimally close matrices $\varrho-\mathrm{d}\varrho$ and 
 $\varrho+\mathrm{d}\varrho$ calculated with respect to the Frobenius norm,  
\begin{equation}
 \label{eq:HSGen}
   \mathrm{g_{{}_\mathrm{HS}}} \propto \mathrm{Tr} \left(\mathrm{d}\varrho\otimes\mathrm{d}\varrho\right)\,.
\end{equation}
If a density matrix belongs to the interior of the simplex $C_{N-1}$\,, i.e., the matrix has $N$ distinct non-zero eigenvalues $(k_1=k_2=\cdots=k_N=1)$\,, then the  metric (\ref{eq:HSGen}) defines the standard \textit{Hilbert-Schmidt ensemble} of random full rank $N\--$qudits. A straightforward computation shows that the joint probability distribution of eigenvalues reads
\begin{equation}
P^{\rm HS}(r_1,\dots,r_N) \propto \,
    \delta(1-\sum_{j=1}^N r_j)  \prod_{j<k}^N (r_j-r_k)^2\,,
\end{equation}
and unitary random factors $U$ in SVD decomposition are  distributed according to the Haar measure on the coset $U(N)/U(1)^N$\,. 


\paragraph{Degenerate Hilbert-Schmidt qudits} If the full rank density matrix has a spectrum of the form (\ref{eq:spec}) with an arbitrary algebraic degeneracy, then the joint probability distribution of eigenvalues is reduced to the following  expression: 
\begin{equation}
P^{\mathrm{HS}}_{k_1, \dots, k_s}(r_1, \dots, r_s) \propto 
 \delta(1- \sum_{i=1}^{s} k_i r_i)   
    \prod_{i<j}^{1\dots s} (r_i-r_j)^{2 k_i k_j}\,.
 \end{equation}
At the same time the angles in the SVD are distributed according to the Haar measure on the coset $U(N)/U(k_1)\times\cdots\times U(k_s)$\,.


\paragraph{Monotone metrics and monotone ensembles of $N\--$qudits}
Two of the above-mentioned metrics, the Bures and Bogoliubov-Kubo-Mori ones, are members of a special class of unitary covariant monotone metrics. According to \cite{Petz1996}, any monotone metric can be written as 
\footnote{
Due to the unitary covariance of the Riemannian metric, it is sufficient to describe monotone metrics evaluated for diagonal  matrices
$D$\,.
}
\begin{equation}
  \mathrm{g}_{{}_D}(X,Y) \propto   \mathrm{Tr}\left(X K^{-1}_D Y \right)\,,
\end{equation}
where $K_D$ is an operator, 
\[
K_D=R_D^{1/2}f(L_D R^{-1}_D) R_D^{1/2}\,,
\]
constructed out of the left and right multiplication operators $L_D$ and $R_D$\,, i.e., $L_DX=DX$
and $R_DX=XD$ for $X \in M_N(\mathbb{C})\,,$
and operator monotone function $f$\,, which  is symmetric, i.e., $f(t)=t f(1/t)$\,, and normalized, $f(1)=1$\,. 

Assuming that $\varrho$ is a full rank density matrix with a simple spectrum and using the 1-form coordinate basis for simplex, $\mathrm{d}r_i\,,$ and non-coordinate basis on $SU(N)$ group, $w_{ij}:=\left(U^\dagger\mathrm{d}U\right)_{ij}$\,, the non-degenerate monotone metrics can be written as
\begin{equation}
 \label{eq:MonotoneM} 
\mathrm{g}_{f}= \frac{1}{4}\sum_{i=1}^N\,
    \frac{\mathrm{d}r_i
    \otimes\mathrm{d}r_i}{r_i}\,  + \frac{1}{2}\,
    \sum_{i< j}^N\,c_f(r_i\,, r_j){(r_i-r_j)^2}\,
    \omega_{ij}\otimes\omega_{ij} \,. 
\end{equation} 
In (\ref{eq:MonotoneM})  by $c_f(x,y) = \frac{1}{y f(x/y)}$  we  denote the Morozova-Chentsov function corresponding to a monotone function $f(t)$\,.
Note that the Bures and Bogoliubov-Kubo-Mori metrics are associated with  the following choice of monotone function:
\begin{equation}
\label{eq:BKM}
f_{{}_\mathrm{B}}(t) =\frac{1+t}{2}\,,
\qquad 
f_{{}_\mathrm{BKM}}(t)= \frac{t-1}{\ln{t}}\,,
\end{equation}
and corresponding Morozova-Chentsov functions,
\begin{equation}
\label{eq:cBW}
c_{{}_\mathrm{B}}(x,y) =\frac2{x+y}\,, 
\qquad
c_{{}_\mathrm{BKM}}(x,y)= \frac{\ln{x}-\ln{y}}{x-y}\,.
\end{equation}


\paragraph{Probability measures from monotone metrics.} 
For an arbitrary monotone metric evaluated for degenerate qudits, calculations of the joint probability distribution of eigenvalues give: 
\begin{equation}
\label{eq:JPDMono}
P^f_{k_1, \dots, k_s}(r_1, \dots, r_s) \propto \frac{\delta\left(1-\sum_{i=1}^{s} k_i r_i\right)}{
\sqrt{r_1\cdot r_2 \cdot \ldots \cdot r_s}} \, 
\prod_{i<j}^s c_{f}^{k_i k_j}(r_i, r_j)\, (r_i-r_j)^{2 k_i k_j}\,.
\end{equation}

\section{Wigner function positivity and classicality }

Here, for the  reader's convenience, before giving a definition of the indicator of classicality, we present the basic settings of the Wigner function of a mixed state of a finite-dimensional quantum system. 
 
\paragraph{Wigner function settings}
The Wigner quasiprobability distribution $W^{(\boldsymbol{\nu})}_\varrho(\Omega_N)$ of an $N\--$level qudit is constructed via dual pairing \cite{AKh2017-WF,AKhT2019}, 
\begin{equation}
\label{eq:WignerFunction}
W^{(\boldsymbol{\nu})}_\varrho(\Omega_N) = \mbox{tr}\left[\varrho
\,\Delta(\Omega_N\,|\,\boldsymbol{\nu})\right]\,,
\end{equation} 
of a density matrix $\varrho$ with the Stratonovich-Weyl (SW) kernel $\Delta(\Omega_N\,|\,\boldsymbol{\nu}) \in \mathfrak{P}^\ast_N\,$  
from the dual space:
\begin{equation}
\label{eq:SWspace}
    \mathfrak{P}^\ast_N=\{\, X \in M_N(\mathbb{C}) \ |\ X=X^\dagger\,,\quad \mbox{tr}\left( X \right) = 1\,, 
    \quad  \mbox{tr}\left( X^2 \right) = N \,\}\,.
\end{equation}
For $N\geq 3$\,, algebraic equations (\ref{eq:SWspace}) admit a family of solutions. As a result, the generic  Wigner function depends on $N-2$ real parameters  $\boldsymbol{\nu}=(\nu_1, \nu_2, \dots, \nu_{N-2} )\,,$ (see details in \cite{AKhT2019}).
The structure of phase space $\Omega_N$  depends on the isotropy group of the SW kernel. For any given isotropy group $H\in U(N)$ of the form 
\[
H={U(k_1)\times U(k_2) \times\ldots\times U(k_{s+1})}\,,
\]
we identify the  phase-space $\Omega_N$ with the  complex flag manifold,  
$$
\Omega_N \,\to\, \mathbb{F}^N_{d_1, d_2, \dots, d_s}= {U(N)} / {H}\,, 
$$
where 
$(d_1, d_2, \dots, d_s)$ is a sequence of positive integers with a sum $N$\,, such that $k_1=d_1$ and $k_{i+1}=d_{i+1}-d_i$ with $d_{s+1}=N\,.$ After presenting necessary notions, we are ready to introduce the  definition of the classicality of states. 


\paragraph{Classical states and classicality indicator} The ``classical states'' form the subset $\mathfrak{P}_N^{(+)} \subset  \mathfrak{P}_N$ of states  whose Wigner function is non-negative everywhere over the phase space: 
\begin{equation}
\label{eq:P+}
   \mathfrak{P}^{(+)}_N = \{\, \varrho \in \mathfrak{P}_N\,\ | \  W_\varrho(z) \geq 0\,, \quad  \forall z\in \Omega_N \, \}\,,
\end{equation}
and similarly, the ``classical states on a fixed stratum'' $ \mathfrak{P}_{H_\alpha}$ are defined as:
\begin{equation}
\label{eq:P+Stratum}
 \mathfrak{P}^{(+)}_{H_\alpha}=
 \mathfrak{P}^{(+)}_N\cap
 \mathfrak{P}_{H_\alpha}\,.
\end{equation}

Based on (\ref{eq:P+}), we define the geometric probability of finding a classical state in an ensemble as 
\begin{equation}
\label{eq:QDef}
\mathcal{Q}_N=
    \frac{\mathrm{Volume (Classical ~ States)}}{\mathrm{Volume(All ~ States)}}\,.
\end{equation}
Here  it is assumed that the Riemannian volume is calculated  with respect to the measure dictated  by the probability  distribution function of an ensemble. Probability (\ref{eq:QDef}) shall be considered as the \textit{global indicator of classicality}. Moreover, for classical states (\ref{eq:P+}) on the fixed stratum $\mathfrak{P}_{H_\alpha}$ we define the \textit{$Q\--$indicator of classicality of the stratum}:
\begin{equation}
\label{eq:QDefStratun}
\mathcal{Q}_N[H_\alpha]=
    \frac{\mathrm{Volume (Classical ~ States~on ~\mathfrak{P}_{[H_\alpha]})}}{\mathrm{
    Volume(All ~States~on~\mathfrak{P}_{[H_\alpha]})}}\,.
\end{equation}
According to (\ref{eq:stratumDeg}), stratum $\mathfrak{P}_{[H_\alpha]}$ consists from subsets of matrices with a certain degeneracy type. Hence, owing to the unitary covariance of probability distribution functions (\ref{eq:InvPDF}), the  $\mathcal{Q}\--$indicator depends only on the joint  probability distribution of eigenvalues of the density matrix and can be rewritten as:
\begin{equation}
\label{eq:QOrb}
\mathcal{Q}_N [H_\alpha]=
\frac{\sum_{\omega \in S_s }\int_{\mathcal{C}^{\ast}_{N-1}(H_\alpha)}
    P^f_{k_\omega(1),\dots,k_\omega(s)}(r_1,\dots,r_s)\,
 \mathrm{d}r_1\wedge\dots\wedge\mathrm{d}r_s}{\sum_{\omega \in S_s }\
\int_{\mathcal{C}_{N-1}(H_\alpha)}
 P^f_{k_\omega(1),\dots,k_\omega(s)}(r_1,\dots,r_s)\,
\mathrm{d}r_1\wedge\dots\wedge\mathrm{d}r_s}\,.  
\end{equation}
In (\ref{eq:QOrb}) the integral in the denominator represents the volume of the orbit space of stratum $\mathfrak{P}_{[H_\alpha]}$\,. The subset $\mathcal{C}_{N-1}(H_\alpha)$ is a union of faces of the simplex $C_{N-1}$ determined by the isotropy group $[H_\alpha]$\,. The integration in the nominator of (\ref{eq:QOrb}) is over the image of $\mathfrak{P}^+_{[H_\alpha]}$  under the canonical quotient map:  
\begin{eqnarray}
\label{eq:OritP+}
\mathcal{C}_{N-1}^\ast(H_\alpha)=  \{\, p(x)\, \ | \ x \in \mathfrak{P}_{H_\alpha}^{(+)}\,\}\,. 
\end{eqnarray}
According to \cite{AKhT2019},  the subset (\ref{eq:OritP+}) can be identified with a certain cone in $\mathbb{R}^{N-1}\,.$
Having denoted by $\boldsymbol{r}=\{r_1, r_2, \dots, r_N\}$ the eigenvalues of the density matrix $\varrho$ and by $\boldsymbol{\pi}=\{\pi_1, \pi_2, \dots, \pi_N\}$  the eigenvalues of the SW kernel, both arranged in decreasing  order, we obtain that $\mathcal{C}_{N-1}^\ast(H_\alpha)$ is the following dual cone:
\begin{equation}
\mathcal{C}_{N-1}^\ast(H_\alpha) =  \
\left\{\, \boldsymbol{\pi} \in \mbox{\bf spec}\left(\Delta(\Omega_N)\right) \ \, |\ \,  ( \boldsymbol{r}^\downarrow, \boldsymbol{\pi}^\uparrow) \geq 0,  \quad \forall\, \boldsymbol{r} \in \mathcal{C}_{N-1}(H_\alpha) \, \right\}\,,
\end{equation}
where $( \boldsymbol{r}^\downarrow, \boldsymbol{\pi}^\uparrow) = r_1\pi_{N}+r_2\pi_{N-1} +\dots +r_{N}\pi_1$\,.

\section{Examples}

In this section results of the  calculations of classicality indicators for the random ensembles described in Section  \ref{sec:Random} will be given for  qubit  and qutrit cases.

\subsection{$\boldsymbol{N=2}\,,$ Qubit}

For a single qubit  the  expansion coefficients of the density matrix over the Pauli $\boldsymbol{\sigma}$\--matrices  are  given by components of the  3-dimensional Bloch vector $\boldsymbol{\xi}=(\xi_1, \xi_2, \xi_3)$: 
\begin{equation}
\label{eq:qubitDM}
    \varrho = \frac{1}{2}\left(I + (\boldsymbol{\xi}, \boldsymbol{\sigma})
    \right)\,.
\end{equation}
Expressing the eigenvalues of (\ref{eq:qubitDM}) in terms of the  length $r\in (0\,, 1]$ of the Bloch vector
\begin{equation}
\label{eq:qubit-r-param}
    r_1 = \frac{1+r}{2}\,, \quad 
    r_2 = \frac{1-r}{2}\,, 
\end{equation}
and taking into account that the Wigner function of a qubit is uniquely constructed with the aid of the SW kernel, whose spectrum is: 
\begin{equation}
\label{eq:qubitSW}
 \pi_1=\frac{1+\sqrt{3}}{2}\,, \qquad \pi_2=\frac{1-\sqrt{3}}{2}\,,
\end{equation}
we conclude that the Wigner function of a qubit is positive definite inside the Bloch ball of radius ${1}/\sqrt{3}\,.$ Since all states, except  the maximally mixed state $r_1=r_2=1/2\,,$  have the torus $T^2\in SU(2)$ as their  isotropy group, there is only one indicator $\mathcal{Q}_{[T^2]}$ for all possible ensembles of qubits. Hence, for any random ensemble of qubits, characterized by a probability distribution $P(r)\,,$ the expression of the classicality indicator is reduced to the following ratio: 
\begin{equation}
\label{eq:Qqubit}
    \mathcal{Q}_{[T^2]}=
    \frac{\displaystyle{\int_{0} ^{\frac{1}{\sqrt{3}}}} P(r) \mathrm{d}r}{\displaystyle{\int_{0}^{1}} P(r) \mathrm{d}r}\,.
\end{equation}


\paragraph{ Hilbert-Schmidt ensemble.} Noting that for the Hilbert-Schmidt ensemble the probability distribution function is $P^{\mathrm{HS}}(r) \propto r^2\,,$ the calculation of (\ref{eq:Qqubit}) gives:
\begin{equation}
\mathcal{Q}^{\mathrm{HS}}_{[T^2]}
= \frac{1}{3\sqrt{3}}\approx 
0.19245\,.
\end{equation} 


\paragraph{Bures ensemble} 
Using the probability distribution function of the Bures ensemble 
\(
P^{\mathrm{B}}(r) \propto \frac{r^2}{\sqrt{1-r^2}}\,,
\)
we find the classicality indicator:
\begin{equation}
\mathcal{Q}^{\mathrm{B}}_{[T^2]}  = \frac{2}{\pi} \left(\arcsin \frac{1}{\sqrt{3}} - \frac{\sqrt{2}}{3}\right) \approx 0.0917211\,.
\end{equation} 


\paragraph{Bogoliubov-Kubo-Mori ensemble } 
Analogously,  calculations with the Bogoliubov-Kubo-Mori measure
$P^{\mathrm{B}}(r) \propto 
\frac{r \left(\ln{\frac{1+r}{2}}-\ln{\frac{1-r}{2}}\right)}{\sqrt{1-r^2}}$
result in the classicality indicator: 
\begin{eqnarray}
\mathcal{Q}^{\mathrm{BKM}}_{[T^2]}  = \frac{2}{\pi} \left(\arcsin \frac{1}{\sqrt{3}} - \sqrt{\frac{2}{3}} \arcoth \sqrt{3} \right) \approx 0.0495506\,.
\end{eqnarray} 

\subsection{$\boldsymbol{N=3}\,,$ Qutrit}
\begin{figure}[h!]
\center{
\includegraphics[width=0.4\textwidth]{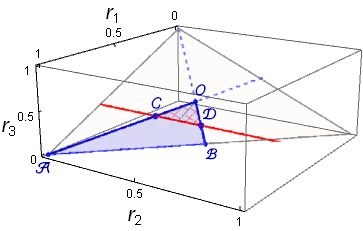}
}
\caption{The ordered 2-simplex of qutrit eigenvalues  is represented by  $\triangle AOB$\,,  and the hatched region, $\triangle COD\,,$   corresponds to the classical states. Edges $AO/\{A\}$  and $BO/\{B\}$ are locus of degenerate  states $F_{12|3}$ and $F_{1|23}$\,, while their parts $CO/\{O\}$  and $DO/\{O\}$ represent the degenerate classical states $F^+_{12|3}$ and $F^+_{1|23}$\,.
}
\label{fig:2S+CL}
\end{figure}

Qutrit state space $\mathfrak{P}_3$ admits the following orbit type decomposition:
\begin{equation}
\label{eq:OTD}
    \mathfrak{P}_3=\mathfrak{P}_{[T^3]}\bigcup
    \mathfrak{P}_{[S(U(2)\times U(1))]}\bigcup
\mathfrak{P}_{[SU(3)]}\,.
\end{equation}
Three strata in (\ref{eq:OTD}) are labeled by the isotropy group or directly by the degeneracy of the density matrices. For full rank states, putting eigenvalues of $\varrho $ in decreasing order, $1 > r_1 > r_2 > r_3 > 0$\,, the components of decomposition  (\ref{eq:OTD}) are described as follows (see geometrical illustration in Fig. \ref{fig:2S+CL}): 
\begin{enumerate}
\item the regular stratum $\mathfrak{P}_{[T^3]}$ of maximal dimension 6   consists  of matrices with a simple spectrum,  $1> r_1\neq r_2\neq r_3 > 0$\,. The corresponding orbit space is the face $F_{123}$ of the ordered 2-simplex, the interior of $\triangle AOB$\,, 
\item the degenerate 4-dimensional stratum $\mathfrak{P}_{[S(U(2)\times U(1))]}$ with density matrices whose degeneracy is $\boldsymbol{k}=(2,1)$ and $\boldsymbol{k}=(1,2)$\,, i.e.,  $1>r_1\neq r_2=r_3> 0$ and $1>r_1=r_2\neq r_3>0$\,. The corresponding orbit space represents the union of edges $F_{1|23}$ and $F_{12|3}$ of the 2-simplex, two sides of $\triangle AOB$\,, 
\item the 0-dimensional stratum $\mathfrak{P}_{[SU(3)]}$ of the maximally mixed state with the triple degeneracy  $\boldsymbol{k}=(3)$\,, $r_1=r_2=r_3=1/3$\,.
\end{enumerate}

Taking into account the decreasing order of the eigenvalues,  $1 \geq r_1 \geq r_2 \geq r_3 \geq 0$\,,  the spectrum of qutrit admits the following  parameterization:
\begin{eqnarray}
\label{eq:specrho1}
r_1&=&
\frac{1}{3}-\frac{2r}{\sqrt{3}}\,
\cos\left(\frac{\varphi+2\pi}{3}\right),\\
\label{eq:specrho2}
r_2 &=&
\frac{1}{3}-\frac{2r}{\sqrt{3}}\,
\cos\left(\frac{\varphi+
4\pi}{3}\right),\\
\label{eq:specrho3}
r_3&=&\frac{1}{3}-
\frac{2r}{\sqrt{3}}\,
\cos\left(\frac{\varphi}{3}\right),
\end{eqnarray}
with  $r \in [0, 1/\sqrt{3}]$ and the angle $\varphi \in [0, \pi]$\,. 
If  $r$ and $\varphi$ are treated as the polar coordinates on a plane, $\left(r\cos\varphi\,, r\sin\varphi\right)\,,$ then  geometrically the formulae (\ref{eq:specrho1})-(\ref{eq:specrho3}) can be interpreted as a map  between  the  ordered simplex $\mathcal{C}_{2}$  and  the domain of the upper half-plane outlined by the Maclaurin trisectrix:
\begin{equation}
    r(\varphi, 1/\sqrt{3})=\frac{1}{2\sqrt{3} \cos({\varphi}/{3})}\,.
\end{equation}

\begin{figure}[h!]
\center{
\includegraphics[width=0.4\textwidth]{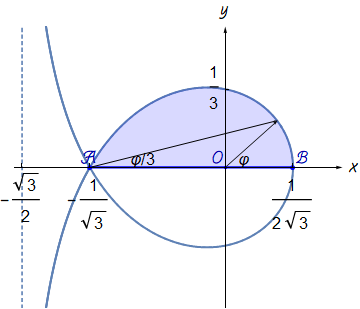} 
}
\caption{The image of the ordered simplex $\mathcal{C}_2$ on the plane $x=r\cos\varphi\,,\, y=r\sin\varphi\,$ under the mapping (\ref{eq:specrho1})-(\ref{eq:specrho3}).}
\label{fig:QutritMacTris}
\end{figure}

More precisely, under  transformations  (\ref{eq:specrho1})-(\ref{eq:specrho3}),  the  ordered simplex of eigenvalues $\mathcal{C}_2$  
maps to the domain (see Fig. \ref{fig:QutritMacTris}) 
\begin{equation}
\label{eq:OrbitQutrit}
 F_{123}=\,: \  
 \biggl\{\, r \geq 0\,,  \varphi \in [0, \pi]\,\ \biggl|\, \ \cos\left(\frac{\varphi}{3}\right) \leq \frac{1}{2\sqrt{3}r} \, \biggl\}\,.
\end{equation} 

\paragraph{Wigner function of a qutrit}
The master equations (\ref{eq:SWspace}) for eigenvalues of the Stratonovich-Weyl kernel of a qutrit,
\begin{equation}
\label{eq:mod3}
\pi_1+\pi_2+\pi_3=1\,,    \qquad  \pi^2_1+\pi^2_2+\pi^2_3=3 \,,  
\end{equation}
define a one-parametric family of the Wigner functions. Due to the permutation symmetry of (\ref{eq:mod3}),
the corresponding moduli space  is  a unit circle factorised by the symmetric group $S_3\,.$ Let $\mu_3$ and $\mu_8$ be  Cartesian coordinates of this arc with a polar angle from the interval $\zeta \in[0, \frac{\pi}{3}]\,,$
 \begin{equation}
\label{eq:muzeta}
\mu_3=\sin\zeta\,, \qquad \mu_8=\cos\zeta\,, 
\qquad 
\end{equation}
then, providing  the decreasing order of the SW kernel eigenvalues, $\pi_1 \geq \pi_2 \geq \pi_3\,,$ one can represent  the whole class of solutions to  (\ref{eq:mod3}) as:    
\begin{eqnarray}
\label{eq:piparam}
\pi_1=  \frac{1}{3}+\frac{2}{\sqrt{3}}\,
\mu_3+\frac{2}{3}\,\mu_8\,,\quad
\pi_2=\frac{1}{3}-\frac{2}{\sqrt{3}}\,\mu_3+\frac{2}{3}\,\mu_8\,, \quad 
\pi_3=\frac{1}{3}-\frac{4}{3}\,\mu_8\,.
\end{eqnarray}


\paragraph{Classical states of qutrit.} 
The image of classical states from the regular stratum $\mathfrak{P}_{[T^3]}$  to the unitary orbit space is the  interior $F^+_{123}$ of a cone which is cut out from  the simplex  $\mathcal{C}_2$  by the line  (see Fig. \ref{fig:2S+CL}) 
\begin{equation}
L_{\boldsymbol{\pi}}(\boldsymbol{r})  \, : \qquad  r_1\pi_3+ r_2\pi_2+ r_3\pi_1=0\,,
\end{equation}
while the orbit space of classical states from the stratum $\mathfrak{P}_{S(U(2)\times U(1))}$ consists of  two pieces: $F^+_{1|23}$ and $F^+_{12|3}$\,, corresponding to the matrices of degeneracy types $(2,1)$ and $(1,2)$ respectively.
Using the polar form of parameterization  of the spectrum of a density matrix 
(\ref{eq:specrho1})-(\ref{eq:specrho3})
and expressions  (\ref{eq:piparam}) 
for  the SW kernel eigenvalues,  the cone of classical states on a regular stratum reads:
\begin{eqnarray}
\label{eq:coneclass1}
&& F^+_{123}\, : \biggl\{\, r > 0\,,  \varphi \in (0, \pi)\,\ \biggl|\, \ \cos\left(\frac{\varphi}{3} +\zeta -\frac{\pi}{3}\right) \leq \frac{1}{4\sqrt{3}r}\,\biggl\}\,,
\end{eqnarray}
while the cone of classical states on the degenerate stratum $\mathfrak{P}^{(+)}_{[S(U(2)\times U(1))]}$ is:
\begin{eqnarray}
   \label{eq:coneclass2} 
   F^+_{1|23}&=&\left\{\,
   \varphi= 0, \ r \in (0, \frac{1}{2\sqrt{3}})\ \biggl| \ 
   \cos\left(\zeta-\frac{\pi}{3}\right)
   < \frac{1}{4\sqrt{3}r}
   \,\right\}\,,\\
   F^+_{12|3}&=& 
   \left\{\,
   \ \varphi= \pi, \ r \in (0, \frac{1}{\sqrt{3}})\ \biggl| \ 
   \cos\left( \zeta\right)
   < \frac{1}{4\sqrt{3}r}
   \,\right\}\,.
\end{eqnarray}

\begin{figure}[h!]
\begin{minipage}[h]{0.32\linewidth}
\center{\includegraphics[width=1\linewidth]{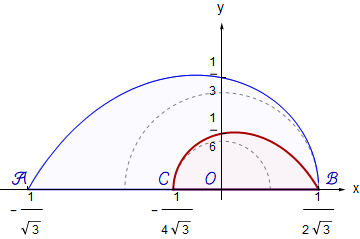}}
\end{minipage}
\hfill
\begin{minipage}[h]{0.32\linewidth}
\center{\includegraphics[width=1\linewidth]{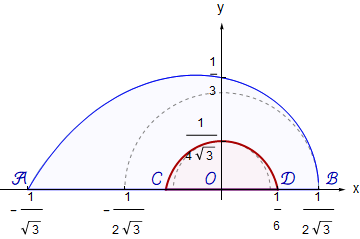}}
\end{minipage}
\hfill
\begin{minipage}[h]{0.32\linewidth}
\center{\includegraphics[width=1\linewidth]{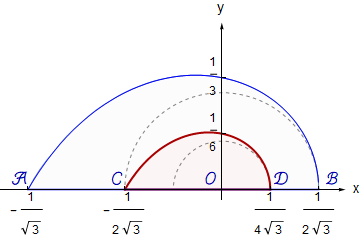}}
\end{minipage}
\begin{minipage}[h]{0.96\linewidth}
\begin{tabular}{p{0.32\linewidth}p{0.32\linewidth}p{0.32\linewidth}}
\centering 
\footnotesize $\zeta=0$ & \centering 
\footnotesize $\zeta=\pi/6$ & \centering 
\footnotesize $\zeta=\pi/3$ \\
\end{tabular}
\end{minipage}
\caption{The orbit space $F_{123}$ in blue and its subspace $F^+_{123}$ in red for different values of the moduli parameter: $\zeta = 0, \pi/6, \pi/3$\,.}
\label{fig:QutritLowerBound}
\end{figure} 

\paragraph{$\mathcal{Q}_3\--$indicator  for Hilbert-Schmidt ensemble of qutrits from regular stratum.} 
The regular stratum $\mathfrak{P}_{[T^3]}$ consists of  density matrices with a simple spectrum.  The expression  $\mathcal{Q}_{[T^3]}$ comprises  the integrals  over the face $F_{123}$ and its subset $F^{+}_{123}$\,:
\begin{equation}
\label{eq:Q3T31}
\mathcal{Q}^{\mathrm{ HS}}_{[T^3]}=\frac{\mathrm{vol_{HS}}(F^+_{123})}{\mathrm{vol_{HS}}(F_{123})}\,.  
\end{equation}
In (\ref{eq:Q3T31})  the expression $\mathrm{vol_{HS}}(X)$   denotes the Riemannian integral over a region $X$ taken with the measure induced on $X \in \mathcal{C}_2$ from the  Hilbert-Schmidt on $\mathfrak{P}_3$\,,
\begin{eqnarray}
\label{eq:volx}
\mathrm{vol}_{\mathrm{HS}}(X)=
\int_{X}\,
P_{1,1,1}^{\mathrm{HS}}(r_1,r_2,r_3)\,\mathrm{d}r_1\wedge\mathrm{d}r_2\wedge\mathrm{d}r_3\,. 
\end{eqnarray}
Taking into account the expression (\ref{eq:HSGen}) for the Hilbert-Schmidt  measure  and the polar form of the parameterization of the qutrit orbit space (\ref{eq:OrbitQutrit}) and of (\ref{eq:coneclass1}),  we obtain  the  indicator of classicality  as a function of the moduli parameter $\zeta$\,: 
\begin{equation}
\label{eq:QR}
\mathcal{Q}^{\mathrm{HS}}_{[T^3]}(\zeta) = \frac{20 \cos^2{\left(\zeta -{\pi }/{6}\right)}+1}{128 \left(4\cos^2{\left(\zeta -{\pi }/{6}\right)} -1\right)^5}\,.
\end{equation}

\begin{figure}[h!]
     \centering
     \begin{subfigure}[b]{0.4\textwidth}
         \centering
         \includegraphics[width=\textwidth]{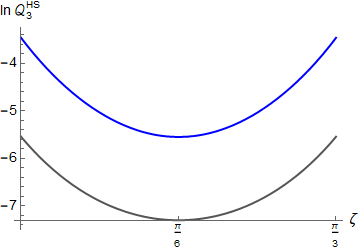}
         \caption{ }
         \label{fig:Qutrit-Q-Ln}
     \end{subfigure}
     \hfill
     \begin{subfigure}[b]{0.4\textwidth}
         \centering
         \includegraphics[width=\textwidth]{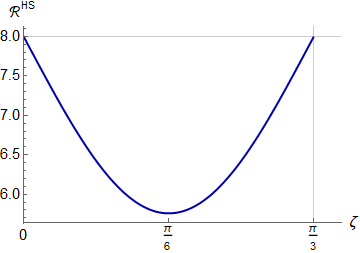}
         \caption{ }
         \label{fig:RatioQregdegenHS}
     \end{subfigure}
\caption{
(a) $\mathcal{Q}_3\--$indicators of a Hilbert-Schmidt qutrit  as functions of $\zeta $ for the regular  (gray curve) and  degenerate (blue curve) strata. The absolute minimum of both indicators is attained at $\zeta=\pi/6$\,. 
(b) The ratio of degenerate to regular  $\mathcal{Q}_3\--$indicators.}
        \label{fig:RHS}
\end{figure}

\paragraph{$\mathcal{Q}_3\--$indicator  for   Hilbert-Schmidt ensemble of qutrits from degenerate stratum.} 
The stratum 
$\mathfrak{P}_{[S(U(2)\times U(1))]}$ has two pieces,  $F_{1|23}$ and $F_{12|3}$, associated with density matrices with degenerate  eigenvalues $r_1=r_2\neq r_3$ and $r_1\neq r_2= r_3$\,, respectively.
Hence,  the $\mathcal{Q}_3\--$indicator for the degenerate stratum of a qutrit reads:
\begin{equation}
\label{eq:Q3U2U1}
\mathcal{Q}^{\mathrm{ HS}}_{[S(U(2)\times U(1))]}=\frac{\mathrm{vol_{HS}}(F^+_{1|23})+\mathrm{vol_{HS}}(F^+_{12|3})}{\mathrm{vol_{HS}}(F_{1|23})+\mathrm{vol_{HS}}(F_{12|3})}\,,  
\end{equation}
where we keep the notation previously used for the regular stratum (\ref{eq:volx}),  noticing only that  the dimension of integration over the degenerate orbit state strata has decreased by one: 
\begin{eqnarray}
\mathrm{vol}_{\mathrm{HS}}(F_{1|23})=
\int_{F_{1|23}}\,
P_{2,1}^{\mathrm{HS}}(r_1,r_2)\,\mathrm{d}r_1\wedge\mathrm{d}r_2\,. 
\end{eqnarray}
The  evaluation of all integrals in (\ref{eq:Q3U2U1}) gives:
\begin{equation}
\label{eq:QD}
   \mathcal{Q}^{\mathrm{ HS}}_{[S(U(2)\times U(1))]}(\zeta) = \frac{1}{1056}\left({\csc ^5\left(\zeta +\frac{\pi }{6}\right)+\sec ^5(\zeta )}\right)\,.
\end{equation}
The functional dependence of the  indicator  $\mathcal{Q}_3^{\mathrm{HS}}$ for the  regular (\ref{eq:QR}) and degenerate  (\ref{eq:QD}) strata is  depicted  in Fig. \ref{fig:Qutrit-Q-Ln}\,. Apart from this,  in Fig. \ref{fig:RatioQregdegenHS} we present the ratio
\begin{equation}
    \mathrm{R^{HS}}(\zeta)=\frac{\mathcal{Q}^{\mathrm{ HS}}_{[S(U(2)\times U(1))]}(\zeta)}{\mathcal{Q}^{\mathrm{ HS}}_{[T^3]}(\zeta)}
\end{equation}
as a certain measure of the relation between the symmetry of a state and its classicality.  

\begin{figure}[h!]
     \centering
     \begin{subfigure}[b]{0.4\textwidth}
         \centering
\includegraphics[width=\textwidth]{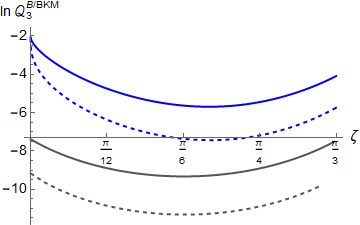}
         \caption{ }
\label{fig:Qutrit-Q-B-BKM-Ln}
     \end{subfigure}
     \hfill
\begin{subfigure}[b]{0.4\textwidth}
\centering
    \includegraphics[width=\textwidth]{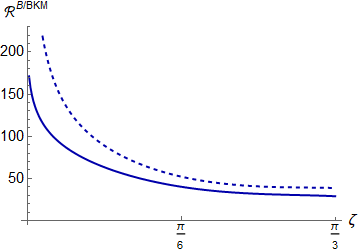}
         \caption{ }
         \label{fig:RatioQregdegenB-BKM}
     \end{subfigure}
\caption{(a) The plot of $\mathcal{Q}_3$ for the  Bures (solid curves) and BKM (dashed curves) ensembles of qutrits from the regular (gray curves) and degenerate (blue curves) strata. 
(b) The ratio $R$ of  degenerate to regular  $\mathcal{Q}_3\--$indicators  for the Bures (solid blue) and the BKM (dashed blue) ensembles.}
\label{fig:2}
\end{figure}

\paragraph{$\mathcal{Q}_3\--$indicator  for   Bures ensemble of qutrits from regular  stratum.}
Using the generic expressions for the joint probability distributions of eigenvalues for monotone metrics (\ref{eq:JPDMono}) and the technique developed above, we compute the $\mathcal{Q}_3\--$indicators  for the Bures and Bogoliubov-Kubo-Mori ensembles of qutrits. The results of our calculations are presented in  Fig. \ref{fig:Qutrit-Q-B-BKM-Ln}.

\section{Summary}

\begin{figure}[h!]
     \centering
     \begin{subfigure}[b]{0.44\textwidth}
         \centering
         \includegraphics[width=\textwidth]{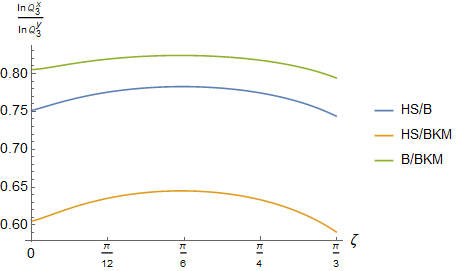}
         \caption{ }
\label{fig:QtrRegRatioslnQlnQ}
     \end{subfigure}
     \hfill
     \begin{subfigure}[b]{0.44\textwidth}
         \centering
         \includegraphics[width=\textwidth]{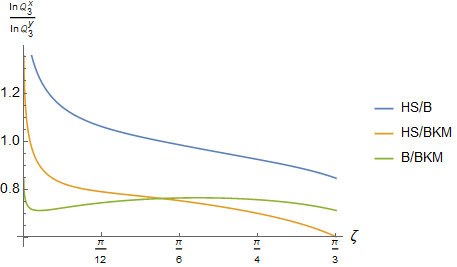}
         \caption{ }
         \label{fig:QtrDegenRatioslnQlnQ}
     \end{subfigure}
\caption{Pairwise ratios  of  $\mathcal{Q}_3\--$indicators of different ensembles for the regular (a) and for the degenerate  (b) stratum.}
\label{fig:PairwiseRat}
\end{figure}

Bearing in mind the results of the calculations of $\mathcal{Q}_3$\,, we will summarize with a few comments. The indicator of classicality  $\mathcal{Q}_3$\,, being a  functional of the ensemble probability distribution function, at the same time depends on two characteristics  of the SW kernel: its isotropy group $H_\alpha$ and the moduli parameter $\zeta$\,. Our studies of the $\mathcal{Q}_3\--$indicator reveal several interesting  peculiarities concerning their interrelations: 
\begin{itemize}
\item There is a certain coherence between the classification of states according to their classicality and their symmetry properties. In particular, it turns out that the states with a ``larger'' symmetry  are more classical, cf. Fig. \ref{fig:RHS} and  Fig. \ref{fig:2}. This observation demands further study and we plan to formalize it in forthcoming publications;
\item The character of the dependence of $\mathcal{Q}_3$ on the type  of the ensemble is monotone, i.e., the values of $\mathcal{Q}_3$ for all strata  are ordered in correspondence with the order of the ensembles, see 
Fig. \ref{fig:PairwiseRat}; 
\item The $\mathcal{Q}_3(\zeta)\--$indicator of the Hilbert-Schmidt ensemble is a symmetric function with respect to the global minimum point, $\zeta=\pi/6$\,, see Fig. \ref{fig:Qutrit-Q-Ln}\,;
\item For monotone metrics the symmetry possessed by the Hilbert-Schmidt ensemble is broken. Data specifying the range of violation is given in Table\,\ref{table:1}.
\end{itemize} 

\begin{table}[h!]
\begin{center}
\begin{tabular}{|p{3cm}|p{3cm}|p{3cm}|p{3cm}|}
\hline\hline
\multicolumn{4}{|c|}{\sc Global $\mathcal{Q}_3\--$indicator vs. moduli parameter}  \\ [0.2ex]
\hline
\hline
{\bf Ensemble }& $\min{\mathcal{Q}_3(\zeta)}$ &$\zeta_{\min}$ & $\mathcal{Q}_3(0)-
\mathcal{Q}_3
({\pi}/{3})$ \\
\hline
Hilbert-Schmidt & 0.0006751 & $\pi/6 \approx 0.523599$ & 0  \\
\hline
BKM  & 0.0000121609 & 0.527798 & 0.0000216102\\
\hline
Bures &
0.0000891011 & 0.525096 & 0.0000472609 \\
\hline 
\end{tabular}
\caption{Data on symmetry properties of $\mathcal{Q}_3\--$indicators.}
\label{table:1}
\end{center}
\end{table}

\section*{Acknowledgments}
The work of A.K. was supported in part by  the Shota Rustaveli National Science Foundation of Georgia, Grant FR-19-034. 


\end{document}